\documentclass[onecolumn,showpacs,preprintnumbers,amsmath,amssymb,aps,letterpaper]{revtex4}
\input{psfig.sty}

\begin{document}
\addtolength{\topmargin}{1.3cm}
%opening
\title{Constraints on the DGP Universe Using Observational Hubble parameter}
\author{Hao-Yi Wan$^1$}
\author{Ze-Long Yi$^1$}
\author{Tong-Jie Zhang$^1$}
\email{tjzhang@bnu.edu.cn}
\author{Jie Zhou$^2$}

%\address{Department of Astronomy, Beijing Normal University,
%Beijing, 100875, P.R.China}

\affiliation{$^1$Department of Astronomy, Beijing Normal University,
Beijing, 100875, P.R.China} \affiliation{$^2$School of Mathematical
Sciences, Beijing Normal University, Beijing, 100875, P.R.China}

\begin{abstract}
In this work, we use observations of the Hubble parameter from the
differential ages of passively evolving galaxies and the recent
detection of the Baryon Acoustic Oscillations (BAO) at $z_1=0.35$ to
constrain the Dvali-Gabadadze-Porrati (DGP) universe. For the case
with a curvature term, we set a prior $h=0.73\pm0.03$ and the best-fit
values suggest a spatially closed Universe. For a flat Universe, we set $h$ free and we
get consistent results with other recent analyses.

\end{abstract}

 \maketitle

\section{Introduction}
Observations of Wilkinson Microwave Anisotropy Probe (WMAP)
\cite{Spergel}, the type Ia Supernova (SN Ia)
\cite{Riess,Perlmutter} and Sloan Digital Sky Survey (SDSS)
\cite{Tegmark,Tegmark+} support an accelerated expanding Universe.
Many cosmological models have been constructed to explain such a
cosmology. Most of them concentrate on the dark energy term with a
negative pressure, within the usual gravitation theory.

The observed accelerated expansion of the Universe is perhaps due to
some unknown physical processes involving modifications of
gravitation theory. Such modifications are usually related to the
possible existence of extra dimensions, giving rise to the so-called
braneworld cosmology. The braneworld cosmology is an example which
excludes the dark energy term by modifying the gravitation theory
\cite{Randall,Randall+,Csaki,Giddings}. One interesting braneworld
cosmological model is the one proposed by Dvali et al., which is
usually called the Dvali-Gabadadze-Porrati (DGP) braneworld
\cite{Dvali,Arkani-Hamed,Dvali+}. For scales below a crossover
radius $r_{\rm c}$, the gravitational force experienced by two
punctual sources is the usual 4-dimensional $1/r^2$ force whereas
for scales larger than $r_{\rm c}$ the gravitational force follows
the 5-dimensional $1/r^3$ behavior.

Although the theoretical consistency and especially its
self-accelerating solution are still waiting for confirming \cite{Luty,Nicolis},
the DGP models have been successfully tested from the
observations. Deffayet et al.
discussed observational constraints from the Cosmic Microwave
Background (CMB) and SN Ia \cite{Deffayet}. Jain et al. presented
a constraint from the viewpoint of gravitational lenses \cite{Jain}.
Alcaniz et al. used the estimated ages of high-$z$ objects to constrain the cosmological
parameters \cite{Alcaniz}. The Chandara measurements of the
X-ray gas mass fraction in galaxy clusters
were used to do a combinational analysis with other cosmological probes \cite{Alcaniz+}.
Pires et al. tested the viability of DGP scenarios from the
cosmological time measurements, i.e., recent estimates of the total age of the Universe
and observations of the lookback
time to galaxy clusters at intermediary and high redshifts \cite{Pires}.
Guo et al. constrained the DGP model from recent supernova observations and BAO \cite{Guo}.
Zhu et al. did the similar work using SN Ia \cite{Zhu}.
%The accelerated expansion Universe can be successfully tested.
See \cite{Lue,Dicus} for
more corresponding comments on the DGP Universe.

In this work, we examine the DGP Universe using the observational
$H(z)$ data (sometimes we call them OHD for simplicity) \cite{Jimenez,Yi}.
The observational $H(z)$ data are related to the differential ages
of the oldest galaxies, the derivative of redshift $z$ with respect
to the cosmic time $t$ (i.e., ${\rm d}z/{\rm d}t$) \cite{Jimenez}.
A determination of ${\rm d}z/{\rm d}t$ provides a measurement of
the Hubble parameter, which can be used as an effective cosmological
probe. In addition, we do the combinational analysis using data of
the size of the Baryonic Acoustic Oscillations (BAO) peak detected in
the large-scale correlation function of luminous red galaxies from
the Sloan Digital Sky Survey (SDSS) \cite{Eisenstein+}. For a
Universe with a curvature term, a prior for the dimensionless Hubble
constant $h=0.73\pm0.03$ is taken from the combinational WMAP
three-year estimate \cite{Spergel}. And we find that the best-fit
values for both two cases suggest a closed Universe. For a flat DGP
Universe, we set $h$ free and get the results consistent with other
independent analyses. The values of the current deceleration parameter
, the transition redshift at which the Universe switches from
deceleration to acceleration and the current
value of the effective equation of state are discussed too.

This paper is organized as follows: In Sec.2, we briefly review the
DGP Universe. In Sec.3, we introduce the observational $H(z)$ data
and the BAO data. In Sec.4, we present the constraints on the
DGP Universe. Discussions and conclusions are given in Sec.5.

\section{Overview of the DGP Universe}

The DGP theory has an important parameter $r_{\rm c}$ which is the
crossover radius where the theory changes between a region that is
effectively 4-dimensional to what is fully 5-dimensional. It is
defined as
\begin{equation}
r_{\rm c}=\frac{M_{\rm Pl}}{2M_5^3},\label{eq1}
\end{equation}
where $M_{\rm Pl}$ is the Planck mass and $M_5$ is the 5-dimensional
reduced Planck mass. In the
DGP Universe, the modified Friedmann equation due to the presence of
an infinite-volume extra dimension reads \cite{Deffayet,Deffayet+}
\begin{equation}
H^2=[\sqrt{\frac{\rho}{3M_{\rm Pl}^2}+\frac{1}{4r_{\rm c}^2}}
+\frac{1}{2r_{\rm c}}]^2-\frac{k}{a(t)^2},\label{eq2}
\end{equation}
where $H$ is the Hubble parameter, $\rho$ is the energy density of
the cosmic fluid and $k=0, \pm 1$ is the spatial curvature
parameter.

If we use the definition
\begin{equation}
\Omega_{\rm{r_c}}=\frac{1}{4r_{\rm c}^2H_0^2},\label{eq3}
\end{equation}
the Hubble parameter can be rewritten as
\begin{equation}
H(z)^2/H_0^2=\Omega_{\rm
k}(1+z)^2+[\sqrt{\Omega_{\rm{r_c}}}+
\sqrt{\Omega_{\rm{r_c}}+\Omega_{\rm m}(1+z)^3}]^2,\label{eq4}
\end{equation}
where $z$ is the redshift, $H_{\rm 0}=100h$ km s$^{-1}$ Mpc$^{-1}$
is the current value of the Hubble parameter, $\Omega_{\rm m}$ and
$\Omega_{\rm k}$ are the matter and curvature density parameters
respectively.
%We plan to consider only the self-accelerating
%phase with $\epsilon=1$ in this work.

And we can get this relation
from the above equation by setting $z=0$,
\begin{equation}
\Omega_{\rm k}+[\sqrt{\Omega_{\rm{r_c}}}+
\sqrt{\Omega_{\rm{r_c}}+\Omega_{\rm m}}]^{2}=1.\label{eq5}
\end{equation}
The current value of the deceleration parameter $q=-\ddot{a}/aH^2$
takes the form \cite{Guo}
\begin{equation}
q_0=(\frac{1}{2}\Omega_{\rm
m}-\Omega_{\rm{r_c}})(\frac{\sqrt{\Omega_{\rm{r_c}}}}{\sqrt{\Omega_{\rm
m}+\Omega_{\rm{r_c}}}}+1)-\sqrt{\Omega_{\rm{r_c}}^2+\Omega_{\rm
m}\Omega_{\rm{r_c}}}.\label{eq6}
\end{equation}
For a flat Universe with $\Omega_{\rm k}=0$, Eq.(\ref{eq5}) reduces
to $\Omega_{\rm{r_c}}=(1-\Omega_{\rm m})^2/4$, from which we get
$0\leq\Omega_{\rm{r_c}}\leq0.25$ for $0\leq\Omega_{\rm m}\leq1$.
The current value of the deceleration parameter can be written as
\begin{equation}
q_0=-1+\frac{3}{2}\frac{\Omega_{\rm m}}{\sqrt{\Omega_{\rm
m}+\Omega_{\rm r_c}}}.\label{eq7}
\end{equation}

If we define $s=\Omega_{\rm r_c}/\Omega_{\rm m}$, the transition
redshift $z_{\rm tr}$ at which the Universe switches from
deceleration to acceleration can be expressed as \cite{Dicus}
\begin{equation}
z_{\rm tr}=-1+2s^{1/3}.\label{eq8}
\end{equation}
Also, we can derive the DGP Universe expressed in Eq.(\ref{eq4})
using the time-dependent effective equation of state \cite{Dicus}
\begin{equation}
\omega_{\rm eff}(z)=-1+
\frac{1}{2}\frac{(1+z)^3}{s+(1+z)^3+\sqrt{s}\sqrt{s+(1+z)^3}}.\label{eq9}
\end{equation}
It is clear that $\omega_{\rm eff}\rightarrow 0.5$ at $z\rightarrow
\infty$. The current value of $\omega_{\rm eff0}$ depends on $s$,
and is always larger than -1.

\section{The Observational $H(z)$ Data Set and BAO}

\subsection{The Observational $H(z)$ Data}

The Hubble parameter $H(z)$ depends on the differential age of the
Universe in this form
\begin{equation}
H(z)=-\frac{1}{1+z}\frac{{\rm d}z}{{\rm d}t},\label{eq10}
\end{equation}
which provides a direct measurement for $H(z)$ through a
determination of ${\rm d}z/{\rm d}t$. By using the differential ages
of passively evolving galaxies determined from the Gemini Deep
Survey Survey (GDDS) \cite{Abraham} and archival data \cite{Treu,
Treu+, Nolan, Nolan+}, Simon et al. determined a set of
observational $H(z)$ data in the range $0\lesssim z\lesssim 1.8$ and used them to
constrain the dark energy potential and its redshift dependence \cite{Simon}.
Using this data set, one can constrain parameters of
various cosmological models. Yi \& Zhang first used them to
analyze the holographic dark energy models in which the
parameter $c$ plays a significant role \cite{Yi}. The
cases with $c=0.6, 1.0, 1.4$ and setting $c$ free are discussed in
detail and the results are consistent with others.
Samushia \& Ratra used this data set to constrain the
$\Lambda$CDM, XCDM and $\phi$CDM models
\cite{Samushia} and Wei \& Zhang analyzed a series of other
cosmological models with interaction between dark matter and dark
energy \cite{Wei}. But as pointed out by Wei \& Zhang, the data
point near $z\sim1.5$ derives from the main trend seriously and dip
down sharply \cite{Wei}. We will omit this point in later discussion.

\subsection{The BAO Data}
The acoustic peaks in the CMB anisotropy power spectrum has been
found efficient to constrain cosmological parameters
\cite{Spergel}.
%Because the acoustic oscillations in the
%relativistic plasma of the early universe will also be imprinted on
%to the late-time power spectrum of the non-relativistic matter
%\cite{Eisenstein}, the acoustic signatures in the large-scale
%clustering of galaxies yield additional tests for cosmology.
Using a large spectroscopic sample of 46,748 luminous red galaxies
covering 3816 square degrees out to $z=0.47$ from the SDSS,
Eisenstein et al. successfully found the peaks, described by
A-parameter that is independent of the dark energy
models\cite{Eisenstein+},
\begin{equation}
A=\frac{\sqrt{\Omega_{\rm
m}}}{z_1}[\frac{z_1}{E(z_1)}\frac{1}{|\Omega_{\rm k}|}{\rm
sinn}^2(\sqrt{|\Omega_{\rm k}|}F(z_1))
]^{1/3},\label{eq11}
\end{equation}
where $E(z)=H(z)/H_0$, $z_1=0.35$ is the redshift at which
the acoustic scale has been measured, the function ${\rm sinn(x)}$
is defined as
\begin{equation}
{\rm sinn(x)}\equiv\left\{
\begin{array}{lll}
{\rm sinh(x)} & {\rm if}\ \Omega_{\rm k}>0;\\
{\rm x} & {\rm if}\ \Omega_{\rm k}=0;\\
{\rm sin(x)} & {\rm if}\ \Omega_{\rm k}<0,
\end{array}\right.\label{eq12}
\end{equation}
and the function $F(z)$ is defined as
\begin{equation}
F(z)\equiv\int_0^z\frac{dz}{E(z)}.\label{eq13}
\end{equation}
Eisenstein et al. suggested the measured value of the A-parameter as
$A=0.469\pm0.017$ \cite{Eisenstein+,Song}. For more information on
BAO, see \cite{Eisenstein+}. The BAO data has been widely used as a
test for cosmological parameters. Wu \& Yu combined BAO with some
recent observational data to determine parameters of a dark energy
model with the equation of state $\omega=\omega_0/[1+b{\rm
ln}(1+z)]^2$ \cite{Wu}. Su et al. combined BAO with GRBs to analyze
the $\Lambda$CDM cosmological model \cite{Su}. It has been declaimed
that BAO is quite robust to constrain cosmological parameters.

More seriously, the BAO data is used only through the fitting
formula given by Eisenstein et al\cite{Eisenstein+}. However, the
A-parameter has been tested only within the limited framework of
standard $\Lambda$CDM (i.e., dark energy models), so it is just
independent of the dark energy models not completely
model-independent. Moreover, the growth of perturbations in the DGP
model is also not the same as that in $\Lambda$CDM \cite{Wayne}, and
therefore the baryon acoustic peak in the DGP model cannot just be
located on the same scale as that in $\Lambda$CDM. Even so, the
discrepancy between the DGP model and dark energy models do not
affect the constraints on the DGP model using BAO data. Guo et al
made the constraints the DGP model using recent supernova
observations and BAO\cite{Guo}, and Pires et al also used BAO to
make a joint statistics for the DGP braneworld cosmology with the
lookback time data set\cite{Pires}.

\section{Constraints on the DGP Universe}

\begin{figure*}
 \centerline{\psfig{figure=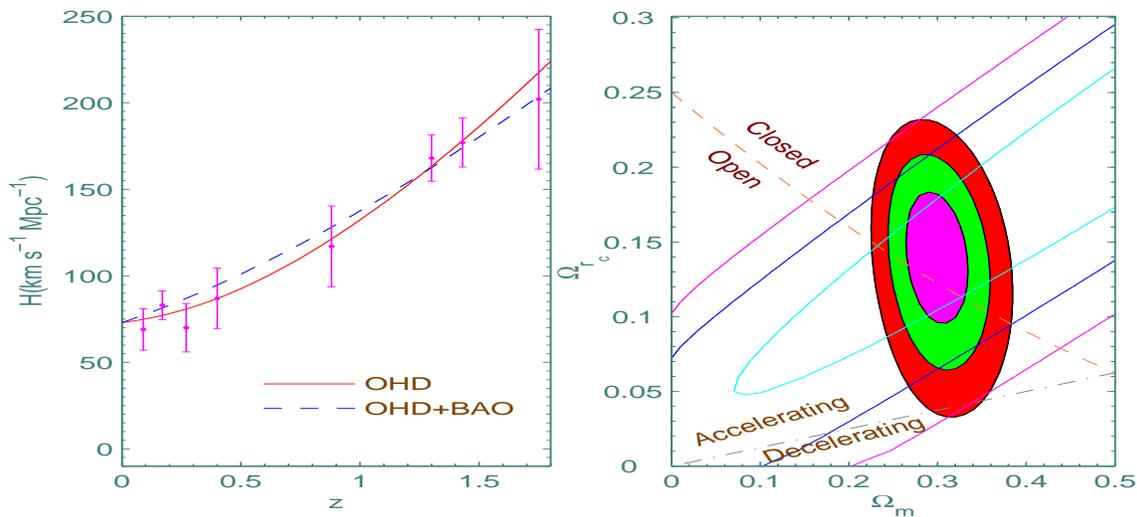,width=6.7truein,height=2.9truein,angle=0}}
{\hskip 0.1in} \caption{Constraints from the observational $H(z)$ data (OHD)
and the BAO data for a non-flat DGP Universe. The left panel: $H(z)$
as a function of $z$ with the best-fit values of $\Omega_{\rm m}$
and $\Omega_{\rm r_c}$, and the observational data with $1\sigma$
error bars are also plotted. The right panel: Confidence regions in
the $\Omega_{\rm m}-\Omega_{\rm r_c}$ plane for the joint analysis
of OHD+BAO (the shaded regions from inner to outer stand for
confidence levels of 68.3\%, 95.4\% and 99.7\% respectively), as
well as analysis of only OHD (the solid lines from inner to outer
stand for confidence regions of 68.3\%, 95.4\% and 99.7\%
respectively).}\label{fig1}
\end{figure*}

First we study the case with a curvature term and assume a prior
of $h=0.73\pm0.03$ from the combinational WMAP three-year estimate
\cite{Spergel}. In order to estimate the best-fit values of
$\{\Omega_{\rm m}, \Omega_{\rm{r_c}}\}$, we use the standard
$\chi^2$ minimization method. If we use only the observational
$H(z)$ data set, we get the fitting results $\Omega_{\rm
m}=0.71\pm0.16$ and $\Omega_{\rm{r_c}}=0.30\pm0.40$. The best-fit
values correspond to a closed and accelerating Universe with
$\Omega_{\rm k}=-1.41$ and $q_0=-0.77$. The current value of the
effective equation of state is $\omega_{\rm eff0}=-0.78$. This
constraint seems very weak due to the large values of the
1$\sigma$ errors and requires combinational analysis with other
cosmological probes. If we combine the observational $H(z)$ data
with BAO, we get $\Omega_{\rm m}=0.30\pm0.02$ and
$\Omega_{\rm{r_c}}=0.14\pm0.03$. The best-fit values suggest a
closed and accelerating Universe too, with $\Omega_{\rm k}=-0.08$
and $q_0=-0.37$. The current value of the effective equation of
state is $\omega_{\rm eff0}=-0.78$. The two cases provide nearly
the same evolutionary values of $\omega_{\rm eff}$. All the
best-fit results are listed in Table\ref{tab1}, as well as $q_0$,
$z_{\rm tr}$ and $\omega_{\rm eff0}$. In the left panel of
Fig.\ref{fig1}, we plot $H(z)$ as a function of $z$ using the
best-fit results for the two cases. And we present the confidence
regions in the $\Omega_{\rm m}-\Omega_{\rm{r_c}}$ plane in the
right panel of Fig.\ref{fig1}, for cases with and without the BAO
data. An accelerating Universe is suggested at 3$\sigma$
confidence level. Larger regions correspond to a closed Universe
even though an open Universe is possible.

\begin{table}
\caption{Fitting results for the corresponding parameters for a
non-flat DGP Universe with a prior $h=0.73\pm0.03$}
\begin{ruledtabular}
\begin{tabular}{lcrlccc}
Test     & $\Omega_{\rm m}$  &  $\Omega_{\rm{r_c}}$  & $r_{\rm
c}$\footnote{in units of
$H_0^{-1}$} & $q_0$  &  $z_{\rm tr}$ & $\omega_{\rm eff0}$\\
\hline
OHD      & 0.71$\pm$0.16               & 0.30$\pm$0.40                 & 0.91 &   -0.77   &  0.50  &    -0.78\\
OHD+BAO  & 0.30$\pm$0.02               & 0.14$\pm$0.03                 & 1.34 &    -0.37  &  0.55  &    -0.78\\
\end{tabular}
\end{ruledtabular}\label{tab1}
\end{table}

\begin{figure*}
 \centerline{\psfig{figure=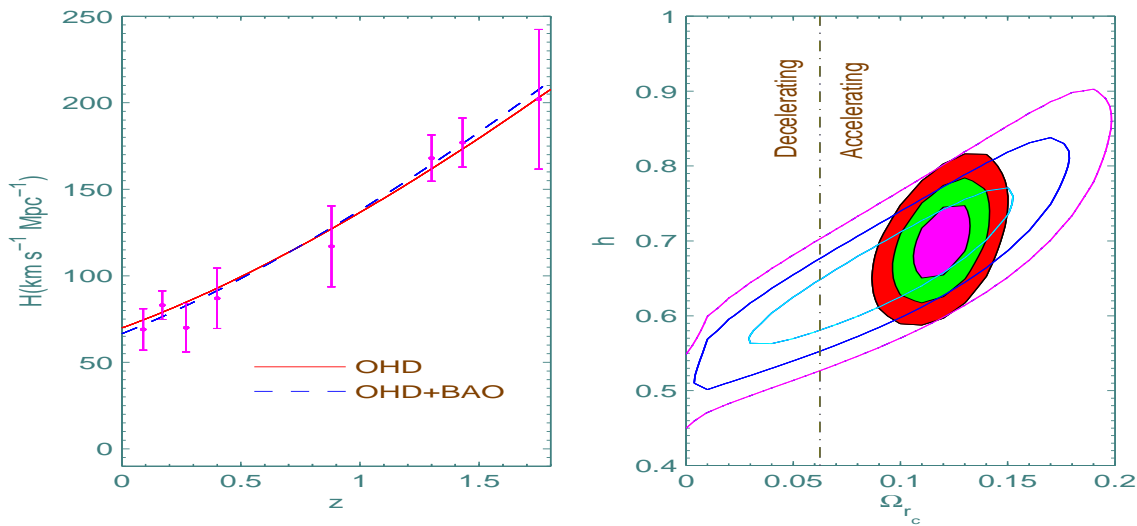,width=6.7truein,height=2.9truein,angle=0}}
{\hskip 0.1in}
 \caption{Constraints from the observational $H(z)$ data (OHD)
and the BAO data for a flat Universe. The left panel: $H(z)$ as
a function of $z$ with the best-fit values of $\Omega_{\rm m}$ and
$\Omega_{\rm{r_c}}$, and the observational data with $1\sigma$ error
bars are also plotted. The right panel: Confidence regions in the
$\Omega_{\rm{r_c}}-h$ plane for the joint analysis of OHD+BAO (the
shaded regions from inner to outer stand for confidence levels of
68.3\%, 95.4\% and 99.7\% respectively), as well as analysis of only
OHD (the solid lines from inner to outer stand for confidence
regions of 68.3\%, 95.4\% and 99.7\% respectively).}\label{fig2}
\end{figure*}

We will look into the flat DGP Universe and we set $h$ free
instead of taking a prior. If we use only the observational $H(z)$
data, we get the fitting results $h=0.67\pm0.07$ and
$\Omega_{\rm{r_c}}=0.10\pm0.04$. The best-fit values correspond to
an accelerating Universe with $q_0=-0.19$. And the current value
of the effective equation of state is $\omega_{\rm eff0}=-0.73$.
If we combine BAO to make a combinational analysis, we get
$h=0.70\pm0.03$ and $\Omega_{\rm{r_c}}=0.12\pm0.09$. The best-fit
values suggest an accelerating Universe with $q_0=-0.29$. And the
current value of the effective equation of state is $\omega_{\rm
eff0}=-0.76$. The values of $\omega_{\rm eff0}$ for the two cases
are close to each other. All the best-fit results are listed in
Table\ref{tab2}, as well as $\Omega_{\rm m}$, $q_0$, $z_{\rm tr}$
and $\omega_{\rm eff0}$. In the left panel of Fig.\ref{fig2}, we
plot $H(z)$ as a function of $z$ using the best-fit results for
the two cases. The confidence regions in the $\Omega_{\rm{r_c}}-h$
plane are presented in the right panel of the same figure, for
cases with and without the BAO data. An accelerating Universe is
suggested at 3$\sigma$ confidence level for combining OHD and BAO.

\begin{table}
\caption{Fitting results for the corresponding parameters for a flat
DGP Universe}
\begin{ruledtabular}
\begin{tabular}{lccrlccc}
Test     & $h$       & $\Omega_{\rm {r_c}}$    & $\Omega_{\rm m}$
&$r_{\rm c}$\footnote{in units of
$H_0^{-1}$}  & $q_0$      &  $z_{\rm tr}$   & $\omega_{\rm eff0}$\\
\hline
OHD      & 0.67$\pm$0.07      &            0.10$\pm$0.04          & 0.37               & 1.58    &  -0.19   & 0.29    &  -0.73\\
OHD+BAO  & 0.70$\pm$0.03      &            0.12$\pm$0.09          & 0.31              & 1.44     &  -0.29   & 0.46    &  -0.76\\
\end{tabular}
\end{ruledtabular}\label{tab2}
\end{table}

\section{Discussions and Conclusions}
Various cosmological observations have been used to explain the
acceleration of the DGP Universe. In this work, we constrain the
cosmological parameters from the observational $H(z)$ data set and
the BAO data. The current values of the deceleration parameter
$q_0$, the transition redshift $z_{\rm tr}$ and the current value of
the effective equation of state $\omega_{\rm eff0}$ have been
derived too. Indeed, the acceleration seems clear for both the cases
with and without a curvature term. For the former case, the best-fit
results correspond to a closed Universe although an open Universe is
possible at larger confidence levels. This is consistent with many
other constraint conclusions \cite{Pires,Riess+,Krisciunas,Astier}.
And the current values of $\omega_{\rm eff0}$ are close to each
other for either combining BAO or not. For the flat Universe, values
of $h$ are a little smaller than the result from the combinational
WMAP three-year estimate \cite{Spergel}. But they are consistent
with the result $h=0.68\pm0.04$ suggested from a media statistics
analysis of the current value of the Hubble parameter
\cite{Gott,Chen}. To make a comparison, we use the same data to do a
constraint on the standard $\Lambda$CDM cosmological model. In
Table\ref{tab3}, we list the values of
$\Delta\chi^2$(DGP-$\Lambda$CDM), which is the excess $\chi^2$ value
between the best-fit DGP Universe and that of $\Lambda$CDM. The DGP
Universe and the $\Lambda$CDM Universe have the same degrees of
freedom no matter whether a curvature term is included. For the DGP
Universe with a curvature term (non-flat), this value is a little
larger than $zero$, which means that the fit for the DGP Universe is
a little poorer than $\Lambda$CDM. For the flat DGP Universe,
$\Lambda$CDM is fit better if only the observational $H(z)$ data are
used, while this is greatly reversed if the two data sets are
considered together. In one word, the observational $H(z)$ data set
can be seen as an acceptable cosmological probe. And the DGP
Universe does not contradict with most observational results. But as
the amount of the observational $H(z)$ data is still so few, there
exist many deficiencies waiting for improving. Combinational
analysis with other observations such as the BAO data is an
efficient way which can provide stronger constraint on the
cosmological parameters.

\begin{table}
\caption{The values of $\Delta\chi^2$(DGP-$\Lambda$CDM)}
\begin{ruledtabular}
\begin{tabular}{lcc}
Test     & non-flat   & flat    \\
\hline
OHD      & 0.006      &     0.123       \\
OHD+BAO  & 0.130      &     -1.744       \\
\end{tabular}
\end{ruledtabular}\label{tab3}
\end{table}

\section{Acknowledgments}

We are very grateful to the anonymous referee for his valuable
comments that greatly improve this paper. This work is supported by
the National Science Foundation of China (Grants No.10473002 and
10533010), the 985 Project and the Scientific Research Foundation
for the Returned Overseas Chinese Scholars, State Education
Ministry.

\end{document}